\documentstyle[multicol,psfig,epsf,aps]{revtex}

\newcommand{\be}{\begin{equation}}
\newcommand{\ee}{\end{equation}}
\newcommand{\bea}{\begin{eqnarray}}
\newcommand{\eea}{\end{eqnarray}}
\newcommand{\ba}{\begin{array}}
\newcommand{\ea}{\end{array}}

\begin{document}
\draft
\title{Comment on ``On two-dimensional magnetohydrodynamic turbulence''
[Phys. Plasmas, 8, 3282 (2001)]}
\author{Mahendra K. Verma  \thanks{e-mail: mkv@iitk.ac.in} \\ 
Department of Physics, 
Indian Institute of Technology, 
Kanpur 208016, India \\
Gaurav Dar  \\
 13, Sukhdev Vihar, New Delhi, India \\
V. Eswaran  \\ 
Department of Mech. Engg.,
Indian Institute of Technology, 
Kanpur 208016, India }

\maketitle

\begin{abstract}
Biskamp and Schwarz [Phys. Plasmas, 8, 3282 (2001)] have reported
that the energy spectrum of two-dimensional magnetohydrodynamic
turbulence is proportional to $k^{-3/2}$, which is a prediction
of Iroshnikov-Kraichnan phenomenology.  In this comment we 
report some earlier results which conclusively show that
for two-dimensional magnetohydrodynamic turbulence,
Kolmogorov-like phenomenology (spectral index 5/3) is better
model than Iroshnikov-Kraichnan phenomenology;  these results
are based on energy flux analysis.
\end{abstract}

\begin{multicols}{2}
In a recent paper, Biskamp and Schwarz \cite{BiskSchw} (now on
referred to as Ref. 1) discuss energy spectrum and structure functions
for two-dimensional (2D) magnetohydrodynamic (MHD) turbulence.  Based
on numerical calculation of energy spectrum, they claim that the
spectrum of 2D MHD turbulence agrees with Iroshnikov-Kraichnan's (IK)
$k^{-3/2}$ law with some modifications.  The purpose of the present
comment is to show that the above claim is inconclusive.  We bring to
notice here an alternative point of view based on energy cascade rates
which supports Kolmogorov's energy spectrum $K^{-5/3}$ for 2D as well
as 3D MHD turbulence; this result is reported in Verma et
al.~\cite{MKV:MHDsimu} and Dar \cite{Dar:thesis}.  

MHD turbulence phenomenologies are discussed
in Verma {\em et al.}~\cite{MKV:MHDsimu} (referred to as Ref.~2).  
For zero velocity-magnetic correlation $({\bf u \cdot b})$, IK
phenomenology predicts
\begin{equation} \label{eqn:Krai65}
E^u(k) = E^b(k) = A (\Pi V_A)^{1/2} k^{-3/2} 
\end{equation}
where $\Pi$ is the total energy flux, $V_A$ is the Alfv\'{e}n
velocity, and $A$ is a universal constant.  Dobrowolny {\em et al.}~\cite{Dobr}
have generalized IK's arguments for nonzero ${\bf u \cdot b}$ and showed
that the energy cascade rates $\Pi^{\pm}$ of ${\bf z^{\pm} = u \pm b}$
are equal irrespective of $E^+$ and $E^-$ ratio, i.e.,
\be
\Pi^+ = \Pi^- \propto \frac{1}{B_0} E^+(k) E^-(k) k^3
\ee
Marsch \cite{Mars:Kolm} 
proposed  Kolmogorov-like phenomenology in which
\be \label{eqn:KolmMHD}
E^{\pm}(k) =  K^{\pm} \frac{\left( \Pi^{\pm} \right)^{4/3}}
 		          {\left( \Pi^{\mp} \right)^{2/3}} k^{-5/3}
\ee
This is a also a limiting case of a generalized phenomenology
of Matthaeus and Zhou \cite{MattZhou}. Clearly,
\be
\frac{E^-(k)}{E^+(k)} = \frac{K^-}{K^+} \left( \frac{\Pi^-}{\Pi^+} \right)^2
\ee

Biskamp and Welter \cite{BiskWelt}, Verma {\em et al.}~\cite{MKV:MHDsimu}, Dar
\cite{Dar:thesis}, and Biskamp and Schwarz \cite{BiskSchw} have
numerically computed the spectral exponents for 2D MHD turbulence.
Biskamp and Welter \cite{BiskWelt} support IK's $k^{-3/2}$ energy
spectra, but Verma {\em et al.}~\cite{MKV:MHDsimu} and Dar \cite{Dar:thesis}
find numerical uncertainties too significant to be able to
distinguish between the exponents 3/2 and 5/3 (see Fig.~1 of Ref.~2).
Biskamp and Schwarz \cite{BiskSchw} do not provide the error bars for
the spectral indices (see Fig.~7 of Ref.~1).  Since 5/3 and 3/2 are so
close, the claims of Biskamp and Schwarz in favor of 3/2 may not be
conclusive.  As stated by them, the intermittency exponents do not
clarify the matter any further.  On the other hand, based on energy
flux studies, Verma {\em et al.}~\cite{MKV:MHDsimu} and Dar \cite{Dar:thesis}
could show quite conclusively that Kolmogorov-like phenomenology
models 2D MHD turbulence better than IK phenomenology.

Verma {\em et al.}~\cite{MKV:MHDsimu} and Dar \cite{Dar:thesis}
numerically computed the energy fluxes $\Pi^+$ and $\Pi^-$ for various
$E^-/E^+$ ratios.  The cascade rates of majority species (larger of
$E^-$ and $E^+$) was always found to be greater that those of minority
species.  To illustrate we have plotted $\Pi^{\pm}$ for $E^-/E^+
\approx 0.2$ in Fig.~1 (taken from Dar \cite{Dar:thesis}).  The same
results are observed in Verma {\em et al.}~\cite{MKV:MHDsimu},
however, the error bars in Dar~\cite{Dar:thesis} is relatively smaller
($\approx 5\%$) because of better averaging.  Clearly $\Pi^+ > \Pi^-$.
Incongruence of the above result with the IK predictions [Eq.~(2)]
clearly indicated that IK phenomenology is not valid for 2D MHD
turbulence.  For $E^-/E^+$ in the range of 0.2 to 1, Verma {\em et
al.}~\cite{MKV:MHDsimu} and Dar \cite{Dar:thesis} find that
\be
\frac{E^-}{E^+} \approx \left( \frac{\Pi^-}{\Pi^+} \right)^2 
\ee 
This result is in agreement with the predictions of Kolmogorov-like
phenomenology for MHD turbulence with $K^+ = K^-$ [Eq.~(4)].
Intermittency corrections to energy fluxes are typically small.
Hence, we do not expect the large difference in energy fluxes to result
from intermittency.  This way Verma {\em et al.}~\cite{MKV:MHDsimu} and
Dar \cite{Dar:thesis} showed that for 2D MHD turbulence,
Kolmogorov-like phenomenology is better model than IK phenomenology.
Recent advances on theoretical fronts also tend to indicate that
Eq.~(4) may even be valid for smaller $E^-/E^+$ with $K^+ \ne K^-$.

Current theoretical and numerical papers \cite{MKV:B0RG,ChoVish:localB} 
argue that Kolmogorov's energy spectrum in MHD is
due to local Alfv\'{e}n effects.  The Alfv\'{e}n waves are
scattered by the ``local mean magnetic field'', rather
than the global mean magnetic field. Hence, the
effective time-scale will be comparable to the nonlocal time
scale resulting in Kolmogorov's energy spectrum for MHD 
turbulence. The above argument is expected to hold in both
2D and 3D.  

To conclude, Verma {\em et al.}~\cite{MKV:MHDsimu} and Dar's
\cite{Dar:thesis} results based on energy fluxes support Kolmogorov's
spectrum for 2D MHD turbulence.  We believe Biskamp and Schwarz's
claim favoring $k^{-3/2}$ energy spectrum is incorrect.

{\em Note added after the receipt of the Reply:} The authors thank
Prof. Biskamp for pointing out Grappin {\em et al.}'s paper
\cite{Grap83}.  However, for high normalized cross helicity ($\approx
0.9$), Verma {\em et al.} \cite{MKV:MHDsimu} and Dar \cite{Dar:thesis}
find the exponents $m^{\pm}$ of $E^{\pm} \approx k^{-m^{\pm}}$ 
in the range of 1.5-1.7, but the ratio
$\Pi^+/\Pi^- \gg 1$ (5 to 10) . Hence numerical results of
Verma {\em et al.}  \cite{MKV:MHDsimu} and Dar \cite{Dar:thesis} are
not in agreement with Grappin {\em et al.'s}
\cite{Grap83} predictions that $\Pi^+/\Pi^- \approx m^+/m^-$.

\end{multicols}

\begin{thebibliography}{10}

\bibitem{BiskSchw}
D. Biskamp and E. Schwarz, Phys. Plasma {\bf 8},  3282  (2001).

\bibitem{MKV:MHDsimu}
M.~K. Verma {\it et~al.}, J. Geophys. Res. {\bf 101},  21619  (1996).

\bibitem{Dar:thesis}
G. Dar, Ph.D. thesis, I. I. T. Kanpur, 2000.

\bibitem{Dobr}
M. Dobrowlny, A. Mangeney, and P. Veltri, Phys. Rev. Lett. {\bf 45},  144
  (1980).

\bibitem{Mars:Kolm}
E. Marsch,  in {\em Reviews in Modern Astronomy}, edited by G. Klare
  (Springer-Verlag, Berlin, 1990), p.\ 43.

\bibitem{MattZhou}
W.~H. Matthaeus and Y. Zhou, Phys. Fluids B {\bf 1},  1929  (1989).

\bibitem{BiskWelt}
D. Biskamp and H. Welter, Phys. Fluids B {\bf 1},  1964  (1989).

\bibitem{MKV:B0RG}
M.~K. Verma, Phys. Plasma {\bf 6},  1455  (1999).

\bibitem{ChoVish:localB}
J. Cho, A. Lazarian, and E.~T. Vishniac, astro-ph/0105235 (accepted in
  Astrophys. J.).

\bibitem{Grap83}
R. Grappin, A. Pouquet, and J. Leorat, Astron. Astrophys. {\bf 126},  51
  (1983).

\end{thebibliography}

\begin{figure}
\centerline{\psfig{figure=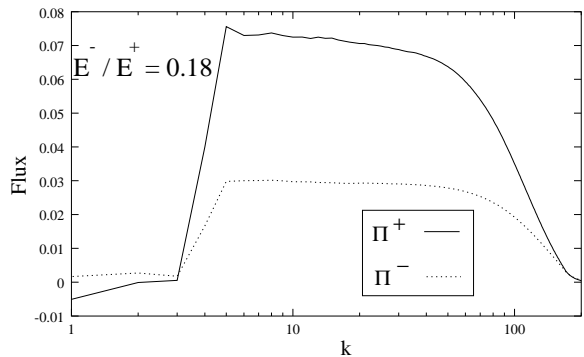,width=3in,angle=0}}
        \vspace*{0.5cm}
\caption{Energy fluxes $\Pi^{\pm}$ versus $k$ for a two-dimensional
run with $B_0 =0.0$ and $E^-/E^+ = 0.18$. Here $\Pi^+ > \Pi^-$.
The figure is taken from Dar [3].}
\label{fig:flux}
\end{figure}

\end{document}